\documentclass[12pt,preprint]{aastex}

\usepackage[cmex10]{amsmath}
\usepackage{rotating}
\usepackage{url}
\usepackage{subfigure}
\usepackage{graphicx}
\usepackage{times}
\usepackage{apjfonts}
\usepackage{isotope}
\usepackage{longtable}
\usepackage{lscape}

\shorttitle{Uranium in the Near Infrared}
\shortauthors{Redman et al.}

\newcommand{\nlines}{$10~100$\space}

\begin{document}

\title{The Infrared Spectrum of Uranium Hollow Cathode Lamps from $850$ nm to $4000$ nm: Wavenumbers and Line Identifications from Fourier Transform Spectra}

\author{Stephen L. Redman\altaffilmark{1,2}, James E. Lawler\altaffilmark{3}, Gillian Nave\altaffilmark{2}, Lawrence W. Ramsey\altaffilmark{1,4}, Suvrath Mahadevan\altaffilmark{1,4}}

\altaffiltext{1}{Department of Astronomy \& Astrophysics, The Pennsylvania State University, University Park, PA 16802}
\altaffiltext{2}{National Institute of Standards and Technology, Gaithersburg, MD 20899}
\altaffiltext{3}{Department of Physics, University of Wisconsin, 1150 University Avenue, Madison, WI 53706}
\altaffiltext{4}{Center for Exoplanets and Habitable Worlds, The Pennsylvania State University, University Park, PA 16802}

\begin{abstract}
We provide new measurements of wavenumbers and line identifications of \nlines UI and UII near-infrared (NIR) emission lines between $2500$ cm$^{-1}$ and $12~000$ cm$^{-1}$ ($4000$ nm to $850$ nm) using archival FTS spectra from the National Solar Observatory (NSO). This line list includes isolated uranium lines in the Y, J, H, K, and L bands ($0.9$ $\mu$m to $1.1$ $\mu$m, $1.2$ $\mu$m to $1.35$ $\mu$m, $1.5$ $\mu$m to $1.65$ $\mu$m, $2.0$ $\mu$m to $2.4$ $\mu$m, and $3.0$ $\mu$m to $4.0$ $\mu$m, respectively), and provides six times as many calibration lines as thorium in the NIR spectral range.  The line lists we provide enable inexpensive, commercially-available uranium hollow-cathode lamps to be used for high-precision wavelength calibration of existing and future high-resolution NIR spectrographs.
\end{abstract}


\section{Introduction}

Astronomical spectrographs are used for a variety of high-precision measurements, ranging from the discovery of low mass exoplanets \citep{2009A&A...507..487M} to the possible variation of fundamental constants, such as the fine structure constant \citep{2001PhRvL..87i1301W} or the proton-electron mass ratio \citep{MalecKeck2010MNRAS}. These works require excellent wavelength calibration sources and a detailed understanding of the associated uncertainties and systematics. In the era of extremely large telescopes (ELTs) it is often the accuracy of the calibration source, not the intrinsic photon noise, that limits the achievable precision. Furthermore, the science goals of future ELTs will require very high-precision calibration sources.  Below $900$ nm, the well-established thorium-argon (Th/Ar) hollow cathode lamps have been a workhorse~\citep{1983ats..book.....P}. Continual improvements in the line list have now enabled Th lamps to be used to calibrate almost the entire optical bandpass with high precision \citep{2007A&A...468.1115L, 2007MNRAS.380..839M}.

In contrast to the optical, the near infrared (NIR) does not yet have a widely-used calibrator like Th/Ar (see \citet{2009ApJ...692.1590M} for a detailed discussion). \citet{kerber2008th} have provided an atlas of Th/Ar lines in the NIR, which is sufficient for precisions of $\approx 50$ m/s on CRIRES (CRyogenic high-resolution InfraRed echelle Spectrograph) \citep{kaeufl2004crires}, but the density of Th lines is relatively low in the NIR, and many of the thorium lines are very weak.  Better precisions of $6$ m/s to $10$ m/s have been achieved by \citet{2010A&A...511A..55F} with Telluric CO$_2$ lines, and $5$ m/s by \citet{2010ApJ...713..410B} with an ammonia (NH$_\textrm{3}$) gas cell in the K band.  Such calibration techniques are suitable for specific problems, but are only useful over specific ranges of the NIR.  

An ideal emission calibration source would cover the observed spectrum in a forest of isolated lines of similar intensity, with the intrinsic wavelength of each line known to a high degree of precision. Laser frequency combs are such a calibration source~\citep{2007MNRAS.380..839M, 2007SPIE.6693E..44O, 2008EPJD...48...57B, quinlan:063105}, but there are numerous technical hurdles before these systems will be considered turn-key: they are expensive to build and operate, they have yet to demonstrate the long-term stability and desired wavelength coverage for the NIR. Hollow cathode lamps, being significantly less expensive and easier to use, are the preferred wavelength calibration solution for most astrophysical spectrographs. We refer the reader to \citet{kerber2007future} for a comprehensive review of the design, construction, and operation of hollow cathode lamps. Thorium ($^{232}$Th), an element often used as the cathode for such lamps, exhibits many of the desired characteristics of an atomic emission calibration source: it has many energy levels (leading to many lines), a heavy nucleus, a very long half-life, and occurs in nature as a single isotope. Uranium shares all of these characteristics except for the last: it has three naturally-occurring isotopes, the major component being uranium-238 (99.275\%), with much smaller amounts of uranium-$235$ ($0.720\%$) and uranium-$234$ ($0.005\%$) \citep{rosman1998isotopic}.  
These isotopes could limit the achievable precision, but the second-most abundant isotope, uranium-$235$, has hyperfine structure, so the fine structure transitions are spread over several lines and are thus much less intense.  For practical purposes, all of the uranium lines are observed as ``single'' lines in the FTS spectra. 
In this paper we propose an inexpensive and simple solution to the calibration problem in the NIR: the use of uranium hollow cathode lamps, which provide a high density of emission lines across the NIR spectral region.

\begin{figure*}[htbp]
\begin{center}
\includegraphics[width=0.48\textwidth,viewport=0 80 612 700, clip]{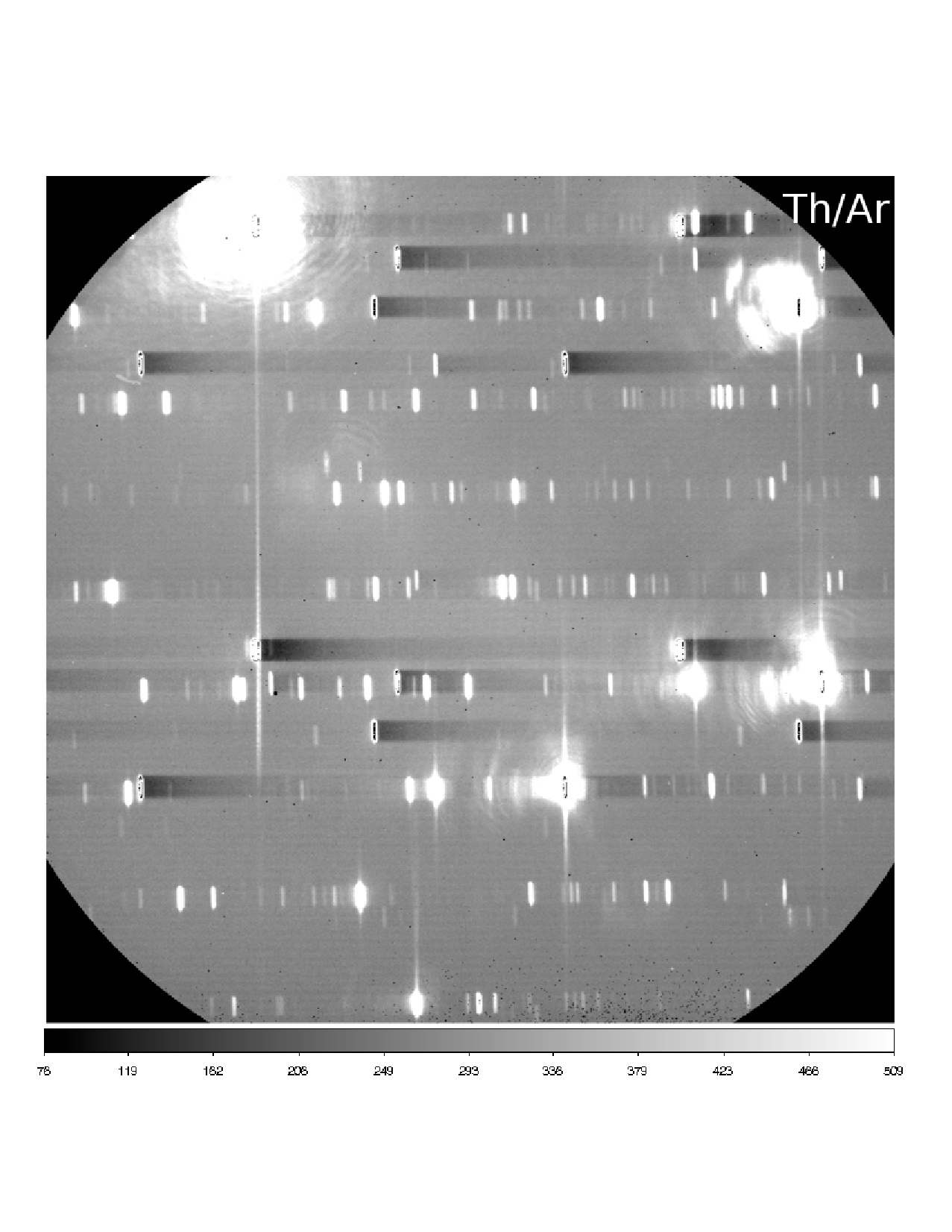}
\includegraphics[width=0.48\textwidth,viewport=0 80 612 700, clip]{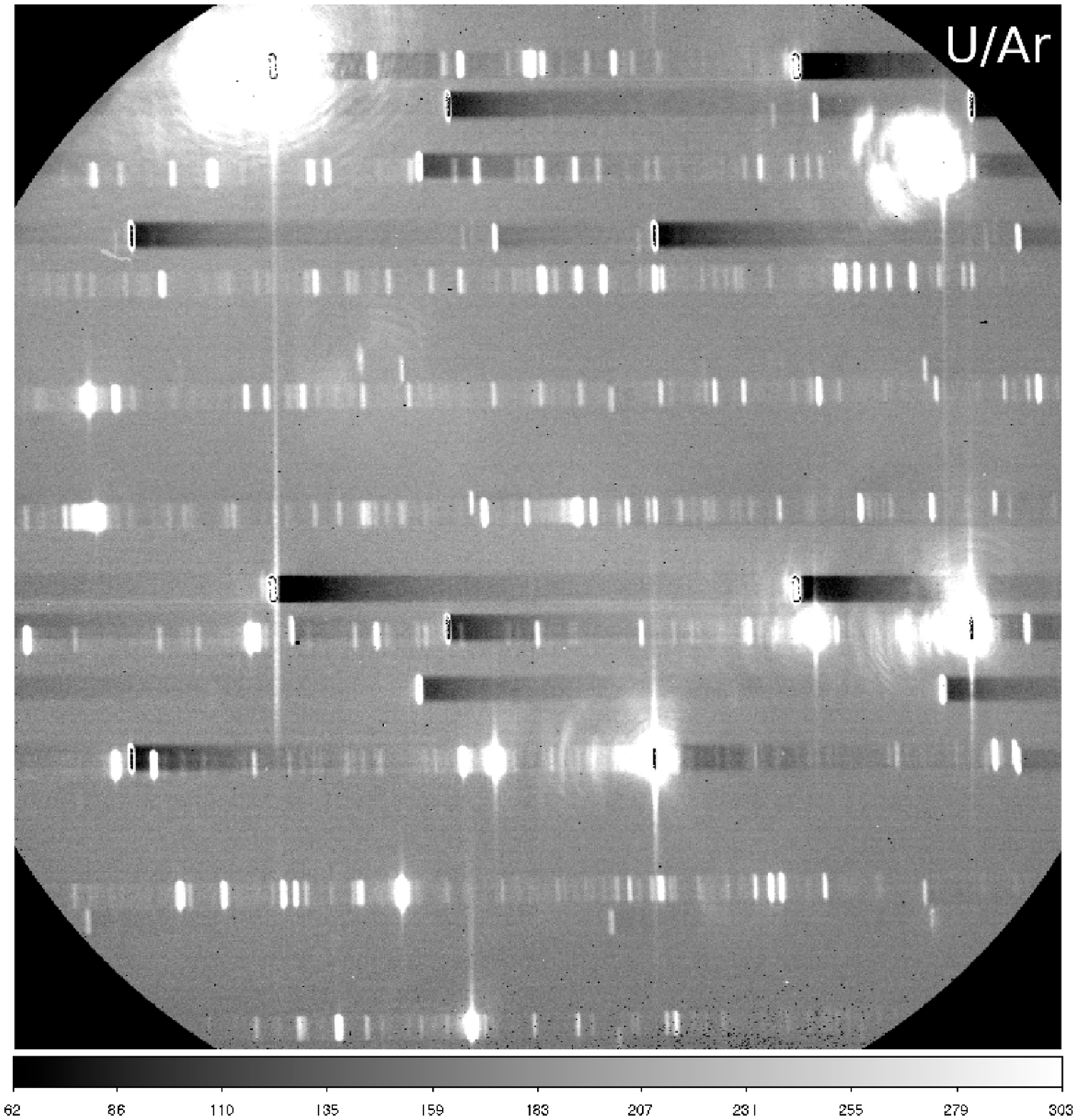}
\includegraphics[width=0.48\textwidth,viewport=0 80 612 700, clip]{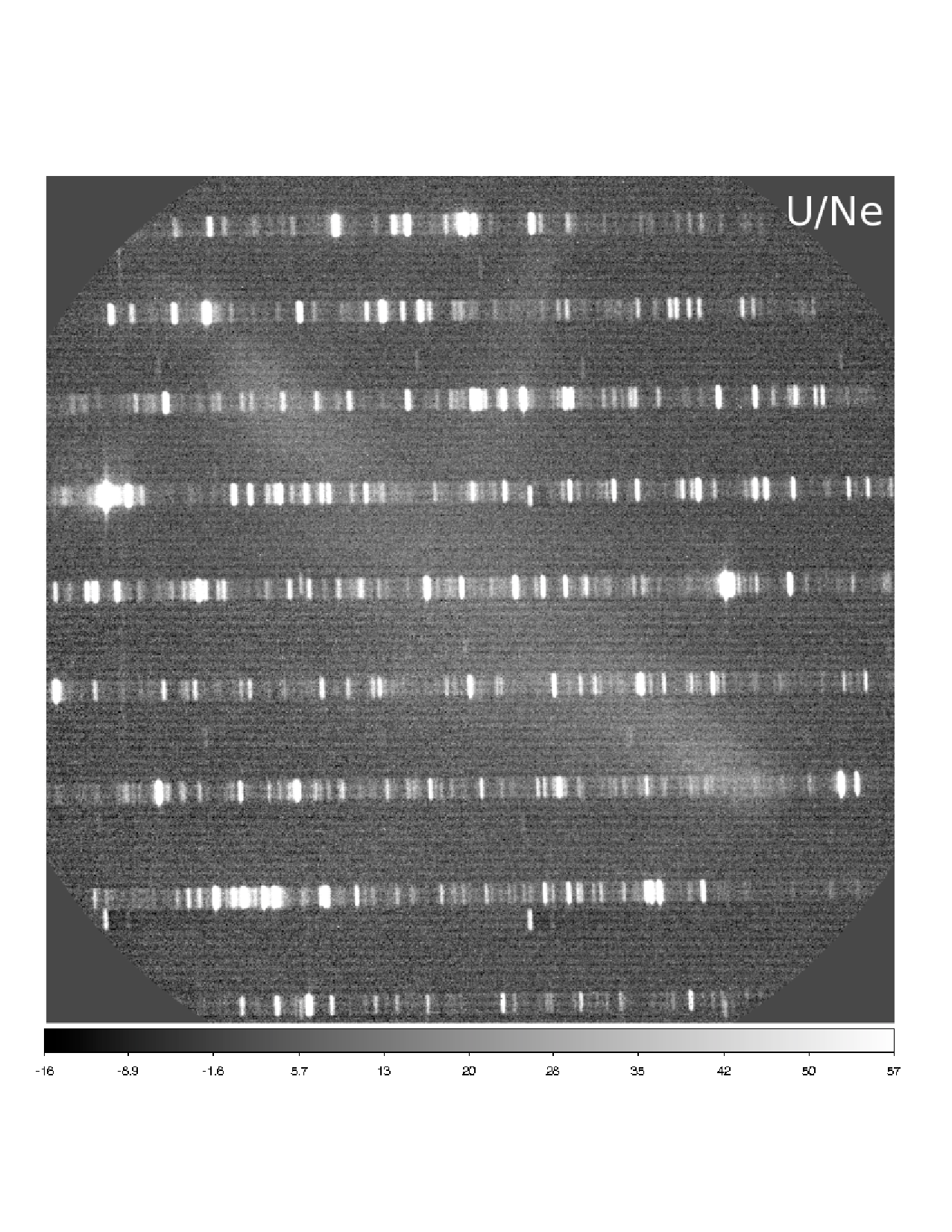}
\caption{Y-band observations with the Pathfinder spectrograph  \citep{ramsey2008pathfinder, 2010SPIE.7735E.231R} of Th/Ar (top-left), U/Ar (top-right) and U/Ne (bottom).  All lamps were run at $14$ mA through the same optical alignment, and all three images are the median-average of twenty-five $25$-second exposures.  Thus, lines common between the top exposures are argon, etc.  The scale of each image is shown at the bottom of each image, and has been adjusted to bring out the faintest lines.  Note the much higher background present in the Ar images; bright argon lines mask out the fainter actinides.  Multiplexer cross-talk is responsible for the negative images of bright lines in adjacent quadrants.  Neon provides a much cleaner image; the faint inter-order scattered light pattern in the U/Ne exposure (visible as a set of faint diagonal streaks) is from three relatively bright neon lines in spectral orders not present on the array.}
\label{fig:comparison}
\end{center}
\end{figure*}

Prior works have noted that uranium might prove a higher line density calibration source than thorium \citep{2003JQSRT..78....1E}. Experiments with a variety of metals and fill gases in hollow cathode lamps on the Pathfinder spectrographic testbed \citep{ramsey2008pathfinder, 2010SPIE.7735E.231R} between $1$ $\mu$m and $1.6$ $\mu$m confirms this. Figure~\ref{fig:comparison} shows $25$-second exposures of Th/Ar, U/Ar, and U/Ne obtained with Pathfinder. All lamps were run at $14$ mA and share the same optical alignment. In addition to the greater line density of uranium lines, we note that neon lines in our U/Ne spectra are much less intense than argon lines in our U/Ar spectra. These results are typical for our observations in the NIR.

Using archival Fourier transform spectrometer (FTS) data from the Kitt Peak National Solar Observatory (hereafter, NSO), we have identified \nlines uranium lines between $0.85$ $\mu$m and $4$ $\mu$m, most of which are below $2$ $\mu$m. This line list complements work by \citet{palmer1980atlas}, who provided uranium identifications for bright uranium lines from $26~000$ cm$^{-1}$ down to $11~000$ cm$^{-1}$ ($3846$ \AA to $9091$ \AA), and \citet{1984ADNDT..31..299C}, who compiled a list of uranium lines between $1.8$ $\mu$m and $5.5$ $\mu$m. Our goal is to provide the astrophysical community with a list of largely unblended wavelength standards that can be used for the calibration of current and future NIR spectrographs.

In \S\ref{sec:data} we describe the archival FTS data from which we derived our measurements and line lists, and in \S\ref{sec:linelist} we describe the construction of the line lists, and the determination of uncertainties. In \S\ref{sec:results} we present the results of this line list, and in \S\ref{sec:discussion} we compare our line list to historical uranium line lists and to recent thorium line lists. In \S\ref{sec:conclusions} we summarize our conclusions.

\section{Fourier-Transform Spectra}
\label{sec:data}

The observations used for our measurements were made with the $1$-m FTS on Kitt Peak  \citep{1976JOSA...66.1081B} between $1979$ and $2002$, and are publicly available on the NSO website\footnote{\url{http://diglib.nso.edu/nso\_user.html}}.  We selected two high-current U/Ar spectra taken with InSb detectors, two low-current U/Ar spectra to check for plasma shifts, and one high-current U/Ne spectrum to search for lines above $9000$ cm$^{-1}$.  Plasma shifts in this paper refer to line shifts from some combination of effects including: collisions with neutral atoms, electrons, and ions, and Stark effect due to static electric fields, and in the case of ion lines the Doppler effect due to ion drift motion.  A summary of these data can be found in Table~\ref{ftsobs}. The first column is the spectrum number (shorthand which is used throughout the text).  The second column is the name of the archive data file (the first six digits indicate the year, month, and day of the observations, respectively, and the last three indicate the spectrum number of that day).  The third column is the lamp type (either uranium neon or uranium argon).  Column four is the lamp current in milliamps (mA).  The wavenumber range of the observation is given in the next two columns.  Column seven is the wavenumber correction factor ($\kappa$, defined below) applied to each spectrum, and column eight is the standard error of the mean in that factor (both divided by $10^{-7}$ for readability).


\begin{table*}
\caption{NSO FTS Observations}
\begin{center}

\begin{tabular}{|c c c c c c c c|} \hline
Spectrum	&	Archive		& Species	& Current 	& \multicolumn{2}{c} {Wavenumber ($cm^{-1}$)}	&	$\kappa$		&	$\delta \kappa$		\\
Number	&	Filename		&		& (mA)	& Start	&	End							&	 ($/10^{-7}$)	&	($/10^{-7}$)		\\ \hline

SP1	&	020227.023	& U/Ar 	& 26 		& 	3900		&	14400 	&	$2.13$		& 	$0.66$		\\
SP2	&	020227.024	& U/Ar 	& 26 		& 	4100		&	14400  	&	$2.09$		& 	$0.64$		\\
SP3	&	790221.005	& U/Ne  	& 75 		&	8300		&	22200 	& 	$-4.62$		&	$0.54$		\\
SP4	&	820520.003	& U/Ar 	& 169	& 	1800		&	9300 	&	$-6.38$		&	$0.83$		\\ 
SP5	&	801214.005	& U/Ar 	& 300 	& 	2400		&	9500 	&	$-15.08$		&	$0.82$		\\ \hline

\end{tabular}
\end{center}
{\bf Notes.} Uranium-argon and uranium-neon FTS spectra from the NSO Archive.
\label{ftsobs}
\end{table*}

These data were reduced and analyzed using the interactive computer program {\sc Xgremlin}, which is an X-windows implementation of Brault's {\sc gremlin} program \citep{brault1989high} that was developed by Griesmann \citep{nave3progress}. The earlier uranium interferograms (SP3, SP4, and SP5) were taken with unequal numbers of points on each side of the zero path difference and the line profiles in the corresponding spectra have large, antisymmteric imaginary parts. Such spectra require careful phase correction in order to ensure that none of the imaginary part of the line profile is rotated into the real part, causing a dispersion-shaped instrumental profile \citep{1995JOSAA..12.2165L}.  We re-transformed SP4 and SP5 from the original interferograms in order to confirm that the phase correction was done correctly. In all cases, the residual phase errors were less than $20$~mrad, corresponding to a maximum wavenumber shift of $0.0003$~cm$^{-1}$.

Independent data reductions of SP5 were carried out by two of us (Redman and Lawler).  In these approaches, lines were kept if they appeared in both SP4 and SP5, had a S/N ratio above $10$, and were symmetric and had a consistent width.  Numerical derivatives (Redman) or integrals (Lawler) were used to determine the center-of-gravity wavenumbers.  In regions where these independent reductions overlapped, matching wavenumbers (within $0.01$ cm$^{-1}$) were averaged with equal weighting.  $97\%$ of the lines identified via these independent analyses were within $0.002$ cm$^{-1}$ of each other.  Where SP4 and SP5 wavenumbers matched within $0.01$ cm$^{-1}$, they were averaged, with twice as much weight given to the higher-current spectrum SP5.

A thermal continuum is observed in our spectra in the wavelength region above 1.5~$\mu$m. This was subtracted from the spectra. All lines with a signal-to-noise ratio of at least $10$ in SP4 and SP5 and $20$ in SP3 were then found in these spectra. Voigt profiles were fitted to these lines to obtain the wavenumber, peak intensity, line width, and Voigt line shape parameters. Since spectrum SP3 was taken in air, all wavenumbers in this spectrum were converted to vacuum wavenumbers with the formulae of \citet{1966Metro...2...71E} using the pressure, temperature, and humidity in the header.  We estimate residual errors in the wavenumbers from this correction to be less than $0.0004$ cm$^{-1}$.  We also calculated these wavenumbers using equation $3$ of \citet{1972JOSA...62..958P}, but found the wavenumbers differed from those obtained with the Edl{\'e}n formula by less than a factor of $4$ parts in $10^9$.

\begin{figure*}[htbp]
\begin{center}
\includegraphics[height=0.8\textwidth,angle=90.]{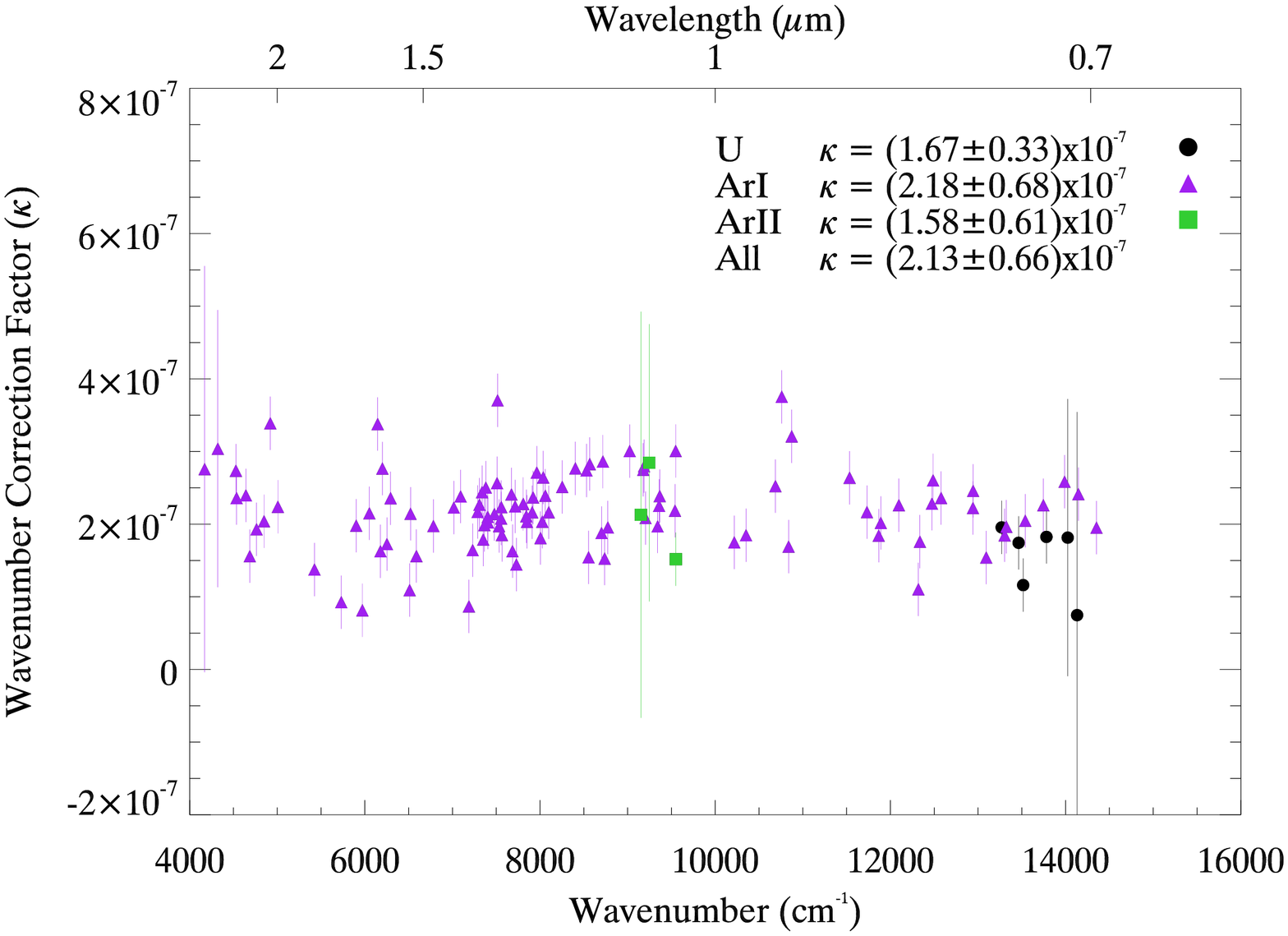}
\includegraphics[height=0.8\textwidth,angle=90.]{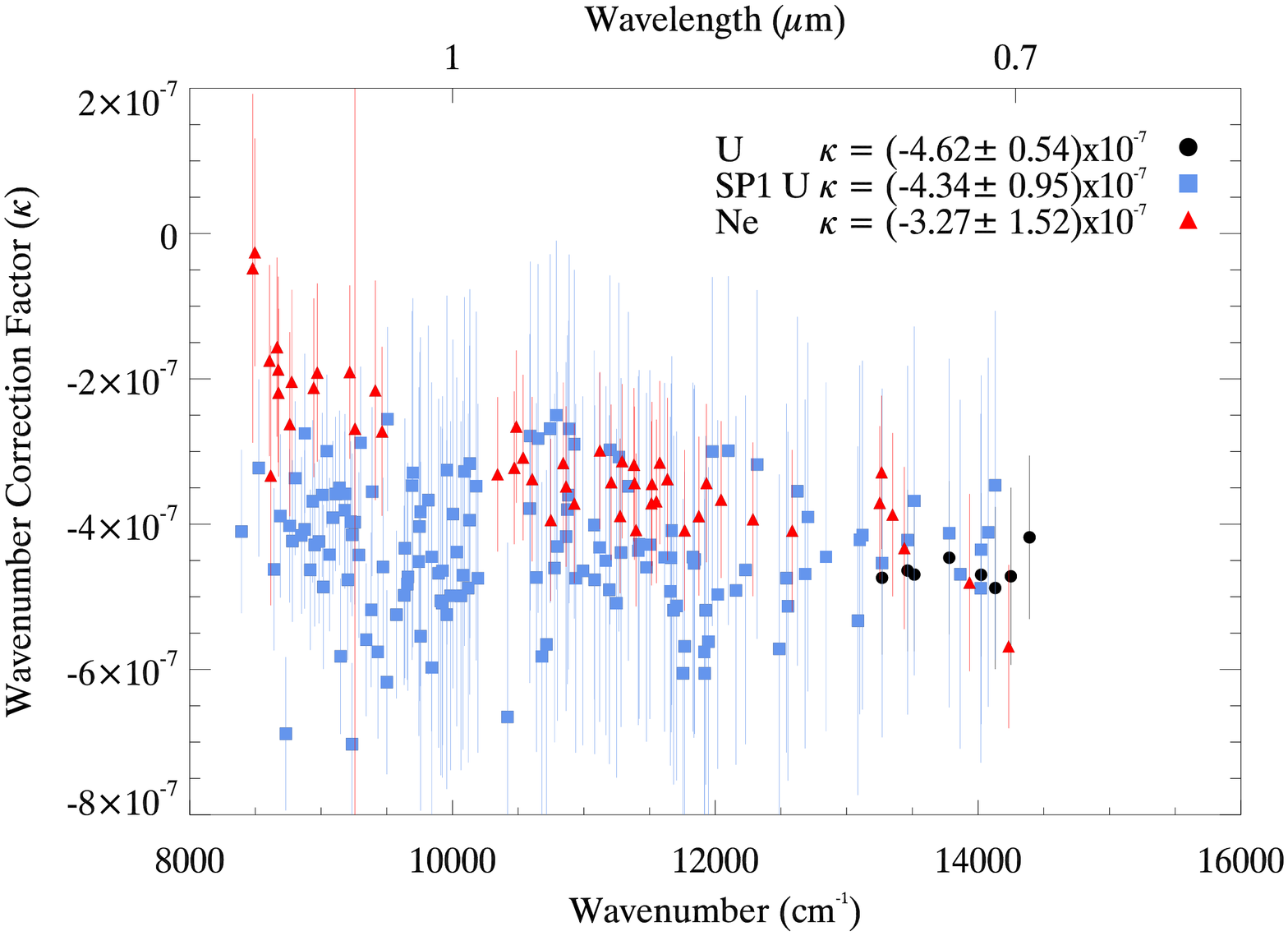}
\caption{{\bf Top:} Wavenumber correction factor of SP1, using wavenumber standards from \citet{whaling2002argon, sansonetti2007comment} (ArI and ArII lines) and \citet{2002JOSAB..19.1711D} (U lines).  Uncertainties in this plot are calculated via Equation~\ref{eq:unc_ind_kappa}.  {\bf Bottom:} The same, but for SP3 (U/Ne).  The neon standards come from \citet{sansonetti2004high}, and a comparison to the uranium lines of SP1 are from this work.}
\label{wncorrfactor_calculation}
\end{center}
\end{figure*}

The wavenumber scale of a FTS is linear and is defined by the control laser used to measure the optical path difference in the interferometer. However, differences in the optical path through the FTS between the laser and the lamp beam result in a small stretching or compression of the wavenumber scale. This effect can be corrected by measuring the wavenumbers of standard lines throughout the spectrum and using them to calculate a multiplicative wavenumber correction factor, $\kappa_i$, using

\begin{equation}
\kappa_{\textrm{i}} = \frac{\sigma_{\textrm{std,i}}}{\sigma_{\textrm{obs,i}}} - 1
\label{eq:kappa}
\end{equation}

\noindent where $\sigma_{\textrm{obs,i}}$ is the wavenumber of a line observed in the spectrum and $\sigma_{\textrm{std,i}}$ is the wavenumber of that line, based on an independent wavelength calibration.  The wavenumber correction factor for the spectrum $\kappa$ is calculated from the weighted mean of the $\kappa_i$s.  All the wavenumbers in the spectrum are multiplied by ($1 + \kappa$) to put them on an absolute scale.  Measurements of $\kappa$ for spectra SP3 and SP5 can be seen in Figure~\ref{wncorrfactor_calculation}.  The legend quotes the trimmed mean (including only points that lie within three standard deviations of the mean $\kappa$) of $\kappa$ for each species, along with an uncertainty that is the sum of the standard error of the mean and the uncertainties of the absolute calibrants.

Wavenumber standards for the calibration of our spectra were taken from three references: Ar~I lines from \citet{whaling2002argon} measured using an FTS, Ne lines from \citet{sansonetti2004high} measured using an FTS, and eight U lines from \citet{2002JOSAB..19.1711D} measured using laser spectroscopy. The Ar~I wavenumbers of \citet{whaling2002argon} have been shown to be too large by $6.7$ parts in $10^8$ by \citet{sansonetti2007comment}. Wavenumbers of Ar~I lines are also susceptible to plasma shifts, but \citet{kerber2008th} showed that these shifts are less than $0.0003$ cm$^{-1}$ ($3$ parts in $10^7$) for Ar~I lines with upper levels less than 115~000 cm$^{-1}$. We thus calibrated spectra SP1, SP2, SP4 and SP5 with Ar~I lines with upper levels below 115~000 cm$^{-1}$ using wavenumbers taken from \citet{whaling2002argon}, with the correction of \citet{sansonetti2007comment}. 

The uranium lines of \citet{2002JOSAB..19.1711D} appear in spectra SP1, SP2 and SP3, but cover a small wavenumber range between 13269~cm$^{-1}$ and 14391~cm$^{-1}$. The calibration of SP1 and SP2 using these lines agrees with the calibration using Ar~I lines to $7$ parts in 10$^8$, which is within the combined uncertainties for spectrum SP2.  Spectrum SP3 contains the uranium lines of \citet{2002JOSAB..19.1711D} and neon lines of \citet{sansonetti2004high}.  The neon standards, however, show a slope in the correction factor of about 2 parts in 10$^7$ between $9000$~cm$^{-1}$ and $14~000$~cm$^{-1}$. Since we observe similar slopes on other uranium/neon spectra we studied during selection of the five spectra we have chosen for this work, we believe this slope is present in the neon lines of \citet{sansonetti2004high} and not in spectrum SP3.  The calibrated uranium lines of SP3 agree with those measured in spectra SP1 and SP2 within the joint uncertainties (See Figure~\ref{wncorrfactor_calculation}).  Therefore, we used only the uranium lines from \citet{2002JOSAB..19.1711D} to calibrate spectrum SP3.

\begin{figure*}[htbp]
\begin{center}
\includegraphics[height=1\textwidth, angle=90.]{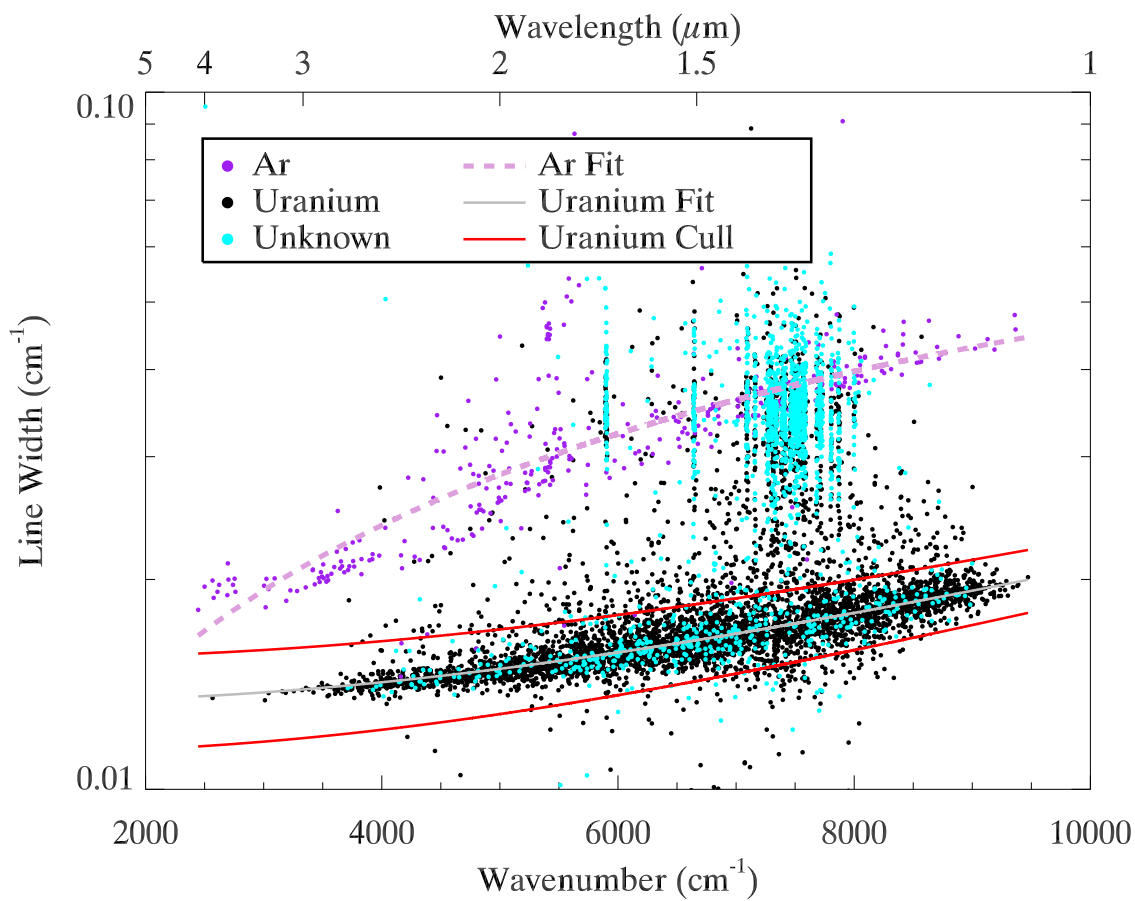}
\caption{A diagnostic plot showing how we first culled blended lines and spurious features from SP5.  Argon lines are wider than uranium lines, and these line widths increase gradually with increasing wavenumber.  These changes are well-fit by a second-order polynomial (gray line for uranium, pink dashed line for argon).  Potential uranium lines with line widths more than three standard deviations from this fit (red lines) were culled from the list.  Approximately $24\%$ of all potential uranium lines were culled in this manner.  This technique also removes blended uranium lines from our line list, which are not as useful for precise wavelength measurements.}
\label{fig:culling}
\end{center}
\end{figure*}

We went through each spectrum individually to check and make sure each feature identified by XGremlin was indeed a line and not a ghost or ringing.  Some of these features were expurgated by modeling the measured uranium line widths (as a function of wavenumber) with a second-order polynomial, and measuring the standard deviation of those line widths.  Potential uranium lines or unidentified lines with line widths greater than or less than three standard deviations from this polynomial fit were culled from the line list.  An example of this is given in Figure~\ref{fig:culling}.  We realize that this technique likely eliminated highly-blended uranium lines, but this is perfectly in line with our goals, since these blended lines cannot be used for precise wavelength calibration.

A relative radiometric calibration of the intensities in Table~2 was performed to convert all intensities to a relative photon flux.  The SP3 and SP5 radiometric calibrations are based on tungsten strip lamp spectra recorded shortly before or after the uranium data and (in the case of SP5) on the relative intensities of ArI lines embedded in the uranium hollow cathode lamp spectra.  The tungsten strip lamp was calibrated  as a spectral radiance standard above $6500$ cm$^{-1}$.  Below this, we estimated the spectrum using a $2400$-K blackbody curve.  We confirmed our calibration by measuring relative intensities of ArI lines in our spectrum and comparing them to published branching ratios \citep{1993JQSRT..50....7W}. During the data analysis we did not observe any self reversed lines.  The extraordinarily rich energy level structure of neutral and ionized U tends to suppress optical depth errors in the relative intensities of Table~\ref{tab:partial}, even at relatively high lamp currents.  

\section{Line List Construction}
\label{sec:linelist}

The quality of spectrum SP5 was noticeably better than spectrum SP3 (which had a lower S/N) and spectrum SP4 (which exhibited spurious ringing around bright lines of an unknown origin), and we therefore elected to weight this spectrum twice as much as these spectra.  This line list was first matched to known argon lines from \citet{whaling2002argon} or neon lines from \citet{sansonetti2004high}, depending on the source, within a $0.01$ cm$^{-1}$ window.  To identify the uranium lines, we first calculated possible transition wavenumbers from energy level differences (Ritz wavenumbers) using the database of actinides\footnote{\url{http://www.lac.u-psud.fr/Database/Tab-energy/Uranium/U-el-dir.html}}, which are based upon \citet{1976JOSA...66..644B} and \citet{1994JOSAB..11.1897B}.  Since these Ritz wavenumbers are based on a least-squares adjustment of all of the lines, the impact of blended lines is minimized.  All lines were compared to these energy level transitions using the same $0.01$ cm$^{-1}$ window function.  Uranium and Noble gas lines were distinguished primarily based upon the line widths.

After this initial identification procedure, roughly 30\% of the lines had more
than one possible classification. Many of these multiple classifications are
incorrect. They occur because the high density of spectral lines and energy
levels in the uranium spectrum gives rise to many spurious coincidences between
measured wavenumbers and energy level differences. The classifications of each of these lines was
examined to determine the most probable classification. The primary criterion was the 
agreement between the observed wavenumber and the wavenumber derived from the 
energy levels (the Ritz wavenumber). For this to be a useful criterion, accurate 
energy levels are needed. The majority of the energy levels from \citet{1976JOSA...66..644B}
and \citet{1994JOSAB..11.1897B} match our spectra within $0.001$ 
cm$^{-1}$. For some levels, the difference between the observed and Ritz wavenumbers 
was much larger. In these cases, a consistent difference was found for
all the lines from the upper level, indicating that the value of the energy level
was not consistent with our data. Although better agreement would be obtained 
by optimizing the energy levels to our data, our spectra do not cover a sufficiently 
large wavenumber region for an optimization of all uranium energy levels. We
have thus used the energy levels from the database of the actinides 
without adjustment. An additional criterion for the acceptance of a 
classification of a line was the energy of the upper level of the transition. 
The majority of the lines in our spectra have upper energy levels below 35~000 cm$^{-1}$ 
and strong lines usually have upper levels at even lower energies. After sorting our line list
in this way, very few ($205$) lines have more than one possible classification.

\subsection{Uncertainties}
\label{ssec:uncertainties}

The uncertainty of a measured line position is a combination of the uncertainty in the measurement and the uncertainty in the standard lines used to set the wavelength scale.  Typically, these two measurements have been treated as if they are independent of each other, so that they are added in quadrature.  However, the uncertainty of the standard lines usually consist of two components: a statistical component, which is the standard deviation of the measurements, and a systematic component, which is common to all of the standard lines (e.g., the standard error of the mean of the original calibration of the standard lines).  The systematic component is usually small, but can accumulate to a non-trivial level if the lines are calibrated with numerous intermediary standards.  In such a case, these measurements are not statistically independent, and should be added together linearly rather than in quadrature.  As such, we have chosen to add our uncertainties linearly.

The weighted uncertainty in the wavenumber correction factor of a particular spectrum ($\delta_{SPX, \kappa}$) is given by:

\begin{equation}
\delta_{SPX, \kappa} = \sqrt{\frac{1}{\sum_{i=1}^{N}{\frac{1}{\delta_{\kappa,i}^2}}}}
+ <\frac{\delta_{\textrm{std,i}}}{\sigma_{\textrm{std,i}}}>
\label{eq:unc_kappa}
\end{equation}

\noindent where the first term is the standard error of the mean of the measured wavenumber correction factors, and the second term is the mean of the ratio of the uncertainty in the wavenumber of the standard, $\delta_{\textrm{std,i}}$, and the wavenumber of the standards, $\sigma_{\textrm{std,i}}$. 
The weighting factor is based on the uncertainty of each of the individual measurements of $\kappa_i$:

\begin{equation}
\delta_{\kappa, i} = \sqrt{ \left(\frac{\partial\kappa}{\partial\sigma_{std,i}}\right)^2 \delta^2_{std,i} + \left(\frac{\partial\kappa}{\partial\sigma_{obs,i}}\right)^2 \delta^2_{lc,i} }
\label{eq:unc_ind_kappa}
\end{equation}

\noindent where $\delta_{std,i}$ is the uncertainty in the line from the literature, and $\delta_{lc,i}$ is the statistical uncertainty of the line center:

\begin{equation}
\delta_{lc,i} = \frac{W_{i}}{S/N_{i}}
\label{eq:unc_stat}
\end{equation}

\noindent where $W_{i}$ is the width of the line, and $S/N_{i}$ is the signal-to-noise ratio of the line.  This equation is derived from equation $9.2$ in  \citet{davis9fourier}, assuming only one statistically-independent point in a line width, which is typical for our under-sampled spectra.

In a given spectrum SPX, the uncertainty in the wavenumber of a line \emph{i} is the sum of the statistical uncertainty (Equation~\ref{eq:unc_stat}) and the uncertainty in the wavenumber correction factor:

\begin{equation}
\delta_{\textrm{SPX,i}} = \sigma_{obs,i} \delta_{SPX,\kappa} + \frac{W_{i}}{S/N_{i}}
\end{equation}

\noindent When a line was found in multiple spectra, these spectrum-specific uncertainties were added in inverse quadrature:

\begin{equation}
\delta_i = \left(\frac{1}{\delta_{SPX,i}^2} + \frac{1}{\delta_{SPY,i}^2} + \cdot\cdot\cdot\right)^{-\frac{1}{2}}
\end{equation}

\section{Results}
\label{sec:results}

The complete table of uranium lines is several hundred pages long, and is available in the electronic version of this publication.  A short selection of the complete list is presented in Table~\ref{tab:partial} to aid in interpretation of the machine-readable (ASCII) table.  

The first column is the weighted mean wavenumber, in cm$^{-1}$, with twice as much weight given to the line position measurements of spectrum SP5 than to spectra SP3 or SP4.  The second column is the combined standard uncertainty in the wavenumber measurement (as outlined in \S~\ref{ssec:uncertainties}, in $10^{-3}$ cm$^{-1}$).  The third and fourth columns are the corresponding vacuum wavelength and wavelength uncertainties.  The fifth column lists the calibrated intensity of the spectral features.  The sixth column provides the species identification, and ionization state for uranium lines.  A ``{\bf ?}'' in this column indicates that the line does not fall within $0.01$ cm$^{-1}$ of a possible uranium transition based on Ritz wavenumbers, but that is most likely uranium, based on the line width.  The next four columns are the upper and lower energy levels for uranium transitions.  The eleventh column is the difference between the Ritz energy level transition and the observed wavenumber of the line.  The last column provides notes on the particular lines; lines that exhibit noticeable asymmetries  are denoted with a {\bf bl} to indicate that they appear to be blended at the resolution of the FTS.


\begin{landscape}
\begin{deluxetable}{lclc rr rcrc rr}

\tabletypesize{\scriptsize}

\tablewidth{0pt}
\tablecaption{Uranium Line List Sample}
\tablehead{
  \colhead{} & \colhead{Wavenumber} & \colhead{} & \colhead{Wavelength} & \colhead{Relative} &
  \colhead{} & \multicolumn{2}{c}{Upper} & \multicolumn{2}{c}{Lower} &
  \colhead{} & \colhead{} \\

  \colhead{Wavenumber} & \colhead{Uncertainty} & \colhead{Wavelength} & \colhead{Uncertainty} & \colhead{Photon Flux} &
  \colhead{Species} & \colhead{Energy} & \colhead{J} & \colhead{Energy} & \colhead{J} &
  \colhead{$\sigma_{Ritz} - < \sigma >$} & \colhead{Notes} \\

  \colhead{(cm$^{-1}$)} & \colhead{($10^{-3}$ cm$^{-1}$)} & \colhead{nm} & \colhead{pm} & \colhead{} &
  \colhead{} & \colhead{(cm$^{-1}$)} & \colhead{} & \colhead{(cm$^{-1}$)} & \colhead{} &
  \colhead{($10^{-3}$ cm$^{-1}$)} & \colhead{}
}

\startdata

 \rule[-1ex]{0pt}{3.5ex}   6467.6270 & 1.1 &   1546.1621 & 0.3 &      3.35 &    ? &                &           &                &           &        &            \\
 \rule[-1ex]{0pt}{3.5ex}   6467.8736 & 0.5 &   1546.1032 & 0.1 &     33.59 &  UII &       $ 15812$ &     $7/2$ &         $9344$ &     $5/2$ &    1.4 &            \\
 \rule[-1ex]{0pt}{3.5ex}   6468.2394 & 0.9 &   1546.0157 & 0.2 &      6.38 &  UII &       $ 19097$ &    $11/2$ &        $12629$ &    $13/2$ &   -1.4 &            \\
 \rule[-1ex]{0pt}{3.5ex}   6468.9299 & 0.6 &   1545.8507 & 0.2 &     86.08 &   UI &       $ 23560$ &       $4$ &        $17091$ &       $3$ &    1.1 &            \\
 \rule[-1ex]{0pt}{3.5ex}   6469.7201 & 1.1 &   1545.6619 & 0.3 &      2.38 &   UI &       $ 27150$ &       $8$ &        $20680$ &       $7$ &   -1.1 &            \\
 \rule[-1ex]{0pt}{3.5ex}   6470.1238 & 0.6 &   1545.5655 & 0.2 &     16.26 &   UI &       $ 24535$ &       $5$ &        $18065$ &       $5$ &    0.2 &            \\
 \rule[-1ex]{0pt}{3.5ex}   6470.2048 & 1.6 &   1545.5461 & 0.4 &      3.32 &  UII &       $ 22868$ &     $9/2$ &        $16397$ &    $11/2$ &    0.2 &            \\
 \rule[-1ex]{0pt}{3.5ex}             &     &           &             &     &   UI &       $ 30443$ &       $2$ &        $23973$ &       $3$ &   -0.8 &            \\
 \rule[-1ex]{0pt}{3.5ex}   6470.2715 & 0.9 &   1545.5302 & 0.2 &      3.87 &   UI &       $ 21329$ &       $7$ &        $14858$ &       $7$ &   -0.5 &            \\
 \rule[-1ex]{0pt}{3.5ex}   6470.9276 & 0.6 &   1545.3735 & 0.1 &     49.91 &   UI &       $ 34086$ &       $5$ &        $27615$ &       $6$ &    3.4 &            \\
 \rule[-1ex]{0pt}{3.5ex}   6471.4221 & 0.6 &   1545.2554 & 0.1 &     16.93 &   UI &       $ 22377$ &       $5$ &        $15906$ &       $6$ &    1.9 &        bl  \\
 \rule[-1ex]{0pt}{3.5ex}   6471.5367 & 0.9 &   1545.2280 & 0.2 &      4.12 &   UI &       $ 18260$ &       $2$ &        $11788$ &       $3$ &    1.3 &            \\
 \rule[-1ex]{0pt}{3.5ex}   6471.6477 & 1.6 &   1545.2015 & 0.4 &      2.72 &  UII &       $ 24923$ &    $13/2$ &        $18451$ &    $13/2$ &    0.3 &            \\
 \rule[-1ex]{0pt}{3.5ex}   6471.8228 & 0.6 &   1545.1597 & 0.1 &     36.65 &   UI &       $ 23932$ &       $5$ &        $17461$ &       $5$ &    0.2 &            \\
 \rule[-1ex]{0pt}{3.5ex}   6472.0924 & 0.6 &   1545.0954 & 0.1 &     64.63 &   Ar &                &           &                &           &        &            \\

\hline

\enddata

\label{tab:partial}

\raggedright
{\bf Notes.}  A small example of the uranium line list.  The classifications are given for UI and UII lines, as are the offsets from the Ritz energy level differences.  Lines unassociated with known uranium level transitions are identified as a ``{\bf ?}''.  These lines are likely also uranium lines, based on their narrow line widths.  A {\bf bl} in the last column indicates a line which appears to be blended at the resolution of the FTS.

\end{deluxetable}
\end{landscape}

Although it is tempting to convert our data to emission branching fractions for various upper levels which decay primarily in the IR, we strongly argue that the reader resist doing so for the following reasons.  First, we did not perform the calibration, and cannot be assured of its validity (although we believe the header data to be accurate).  Second, our intensity calibration is less reliable below $6500$ cm$^{-1}$, where we used a $2400$-K blackbody curve and ArI branching fractions to estimate the tungsten strip spectrum beyond its measured range.  Third, any branching fractions deduced solely from this work will be incomplete, because we only analyze the IR.  At the very least, some additional tests are desirable.  Specifically, a check for possible optical depth errors on some of the strongest emission lines to low lying levels should be performed by re-measuring their branching ratios over a range of lamp currents.  

\begin{figure*}[htbp]
\begin{center}
\includegraphics[height=1\textwidth,angle=90.]{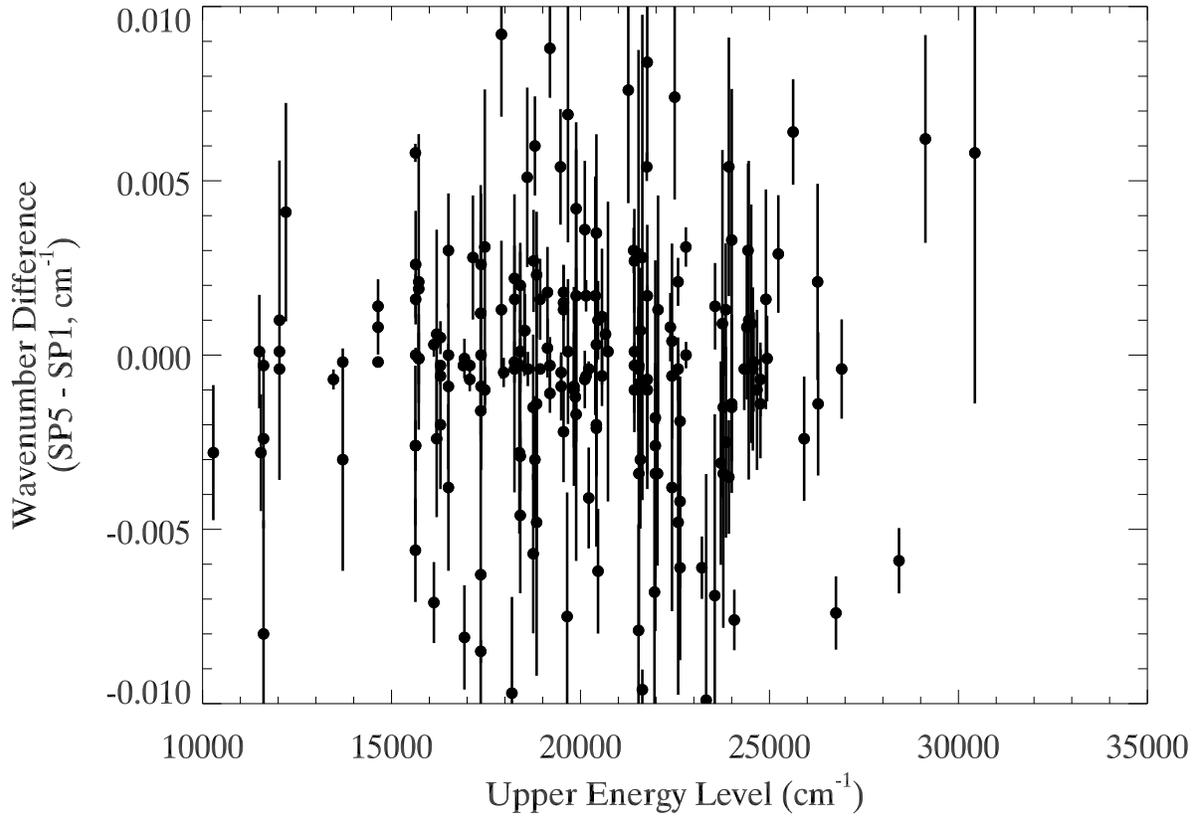}
\caption{The wavenumber difference (in cm$^{-1}$) between our highest-current spectrum (SP5, $300$ mA) and our lowest-current spectrum (SP1, $26$ mA) as a function of the upper energy level of the transition.  The uncertainties are the sum in quadrature of the uncertainties in SP1 and SP5.  The lack of a trend indicates that plasma shifts are not an issue for these upper energy levels.}
\label{fig:plasma_shifts}
\end{center}
\end{figure*}

The physical conditions of the plasma inside the hollow cathode lamp can shift the spectral lines systematically.  Such shifts are found by comparing data taken under a variety of conditions.  We compared the lines from our low-current observations (SP1 \& SP2) to the lines of our highest-current observation (SP5).  No systematic plasma shifts were observed at the $10^{-3}$ cm$^{-1}$ level (see Figure~\ref{fig:plasma_shifts}).

In all, we identified $8991$ UI lines and $1150$ UII lines.  $1244$ lines remain unidentified, but are likely to be UI or UII, based on their narrow line widths.   We see no uranium-$235$ lines in our spectra, both because of the relatively low abundance of this isotope, and because hyperfine structure lowers the intensity of these lines relative to uranium-$238$.

To assist with the use of this line list in the H band, we have constructed an atlas of uranium-neon lines between $14~540$ \AA~and $16~380$ \AA.  This atlas was measured at a resolution of $50~000$ using Pathfinder, and independently calibrated with a laser frequency comb.  The U/Ne hollow cathode lamp was run at $14$ mA.  Measurements of the spectral peaks of these lines show excellent agreement with the FTS-measured wavenumbers in this publication.  This spectroscopic atlas will be published separately \citep{RedmanAtlas2011InPrep}.

\section{Comparison with Other Uranium and Thorium Line Lists}
\label{sec:discussion}

\begin{figure*}[htbp]
\begin{center}
\includegraphics[height=1\textwidth,angle=90.]{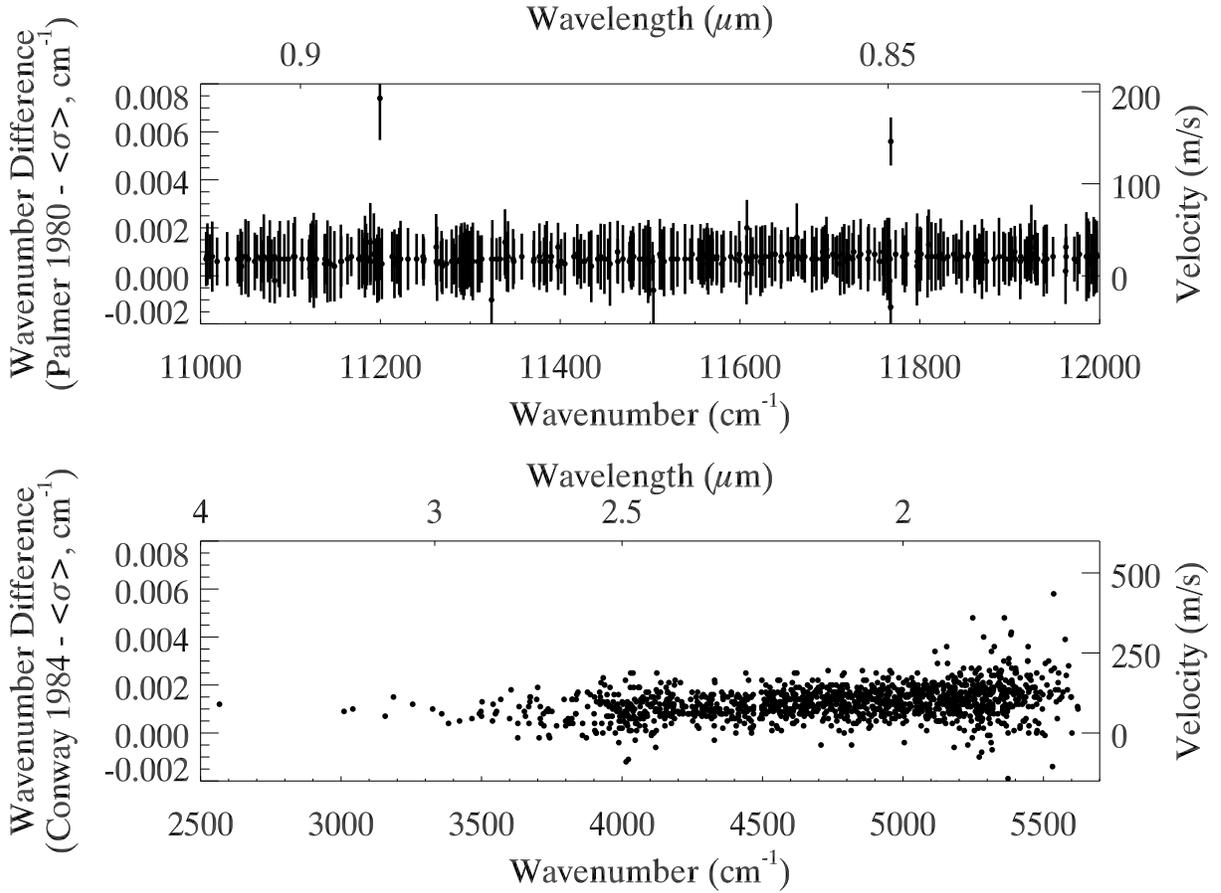}
\caption{Comparisons to historical uranium line lists.  {\bf Top:} A comparison to the lines of \citet{palmer1980atlas}.  The average offset and one standard deviation of this offset is ($0.75\pm0.53$)$\times10^{-3}$ cm$^{-1}$.  The equivalent velocity offset and single standard deviation (using the central wavenumber) is ($20\pm14$) m/s.  The fact that the statistical deviation between the two data sets is smaller than the estimated uncertainty for most lines indicates that \citet{palmer1980atlas} did a good job estimating the uncertainty of their uncalibrated data.  Some outliers are weak lines near bright, unresolved uranium lines (which exhibit ringing); these lines are difficult to measure, and our measurements differ slightly from those of \citet{palmer1980atlas}. {\bf Bottom:} A comparison to the lines of \citet{1984ADNDT..31..299C}. The average offset and one standard deviation of this offset is ($1.26\pm0.80$)$\times10^{-3}$ cm$^{-1}$.  Error bars in each are the sum in quadrature of our uncertainties and the uncertainties of the aforementioned authors.  In the case of \citet{1984ADNDT..31..299C}, we have omitted the error bars for readability; the average error bar in this case is $\pm 0.001$ cm$^{-1}$.  The equivalent velocity offset and single standard deviation (using the central wavenumber) is ($57\pm36$) m/s.}
\label{fig:compare_p80_c84}
\end{center}
\end{figure*}

Our line list overlaps those of \citet{palmer1980atlas} between $11~000$ cm$^{-1}$ and $12~000$ cm$^{-1}$, and of \citet{1984ADNDT..31..299C} between $2500$ cm$^{-1}$ and $5500$ cm$^{-1}$. A comparison of our wavenumbers is shown in Figure~\ref{fig:compare_p80_c84}. All the wavenumbers of Palmer \& Engleman are ($0.75\pm0.53$)$\times10^{-3}$ cm$^{-1}$ greater than ours. \citet{palmer1980atlas} did not calibrate their spectra with standard lines and the only corrections they made to their wavenumber scale were due to the refractive index of air and the finite aperture of the FTS. This likely accounts for the difference between our two wavenumber scales.  That being said, their estimated uncertainties overlap our measurements within one or two standard deviations.  Several outliers are low-S/N lines near bright lines, and as such are difficult to measure.  Our measurements of these lines differ slightly from those of \citet{palmer1980atlas}, but these lines are likely to not be very useful for high-precision calibration anyway.

The comparison to \citet{1984ADNDT..31..299C} (bottom frame of Fig.~\ref{fig:compare_p80_c84}) is on the same scale and shows a much larger scatter.  This is not surprising, considering their measurements have an order of magnitude less precision than our own measurements.  Even so, our measurements agree to within ($1.26\pm0.80$)$\times10^{-3}$ cm$^{-1}$.

\begin{figure*}[htbp]
\begin{center}
\includegraphics[height=1\textwidth,angle=90.]{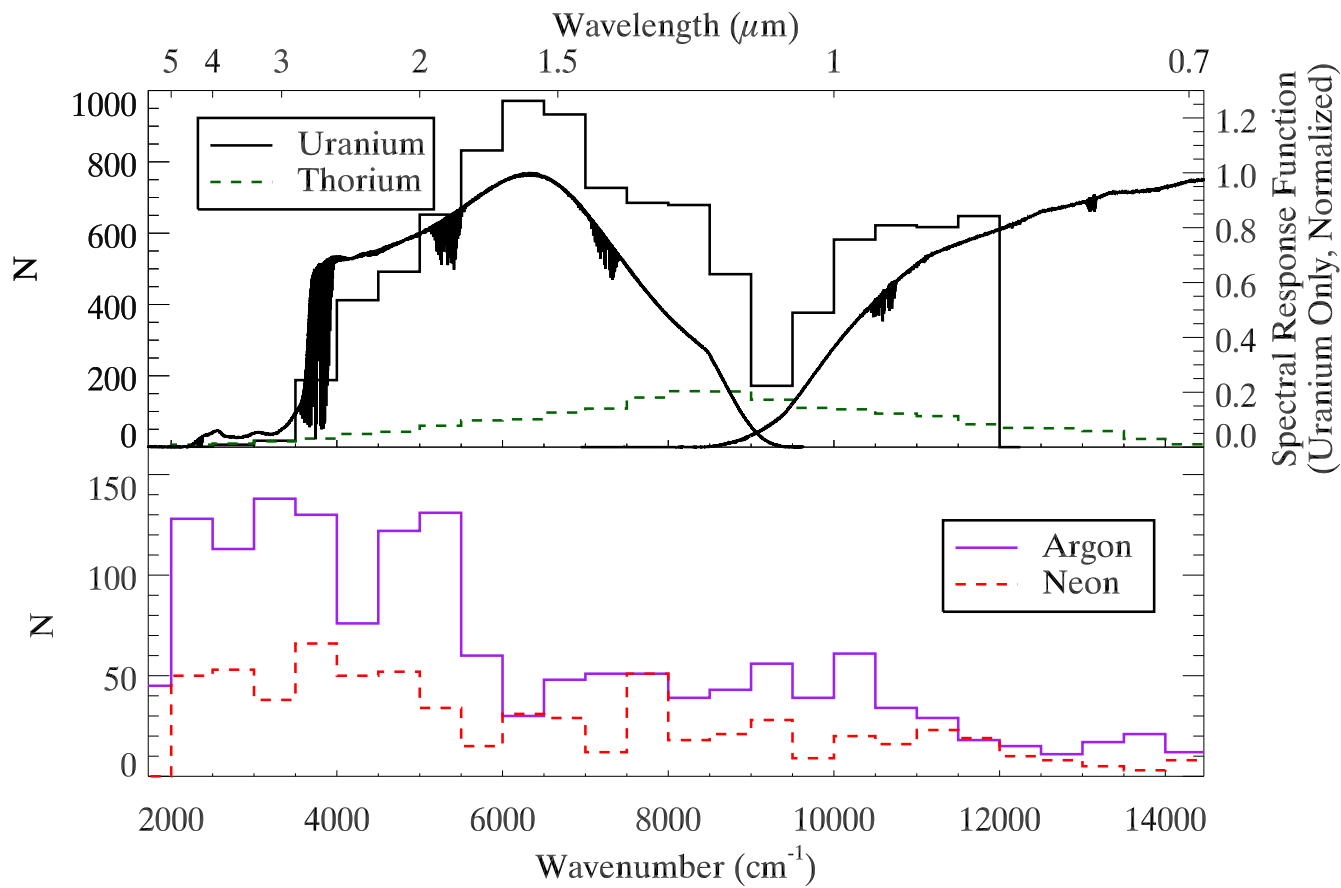}
\caption{Histograms of the number of documented standard uranium, thorium, argon, and neon lines in the NIR.  It is important to note that these distributions are not necessarily proportional to the intrinsic distribution density of these spectral lines, since these lines are compiled from various sources and different observation conditions.  {\bf Top:} Histograms of the uranium (this work) and thorium \citep{kerber2008th} lines in the NIR.  The large gap in the number of uranium lines around $9250$ cm$^{-1}$ is from a gap in the spectral response functions of these archived data.  The spectral response function of SP5 (left) and SP3 (right) are based upon tungsten strip lamp spectra taken with the same observing conditions as the U/Ar or U/Ne spectra.  The drop in the number of uranium lines above $12000$ cm$^{-1}$ is purely artificial, based on our chosen upper-limit for this compilation.  {\bf Bottom:} The same, but for argon \citep{whaling2002argon, sansonetti2007comment} and neon standards \citep{sansonetti2004high}.}
\label{fig:fig_y}
\end{center}
\end{figure*}

The top plot in Figure~\ref{fig:fig_y} shows a histogram comparing the population density of uranium lines from this work and thorium lines from \citet{kerber2008th}.  Note that the lamp used in the latter was run at $20$ mA, much less than the lamp currents of the FTS uranium observations, so more thorium lines undoubtedly exist in the NIR that have not yet been published.  The difference in quantity between this line list and Kerber's excellent Th/Ar work is quantified in Table~\ref{tab:kerber}.  The gap in the uranium line density around $9250$ cm$^{-1}$ is an artifact of the spectral response function of the archival data used to create this line list, and not representative of the true number of uranium lines in this region.  The response functions of spectra SP5 and SP3 are plotted in this same panel, and are based upon tungsten strip lamp spectra taken under the same observing conditions as their corresponding uranium spectra.

The bottom plot in Figure~\ref{fig:fig_y} shows a comparison between the number of neon and argon lines in the same spectral region (bottom).  As shown in Fig.~\ref{fig:comparison}, not only are the neon lines less numerous, they are also much fainter than argon lines in the same spectral region.  


\begin{table}
\caption{Thorium and Uranium Line Quantities in Astrophysical Bands of the NIR}
\begin{center}
\begin{tabular}{|c |c |c |c|} \hline
Band&	Wavelength 		&	N$_{Th}$	&	N$_{U}$	\\
	&	Range ($\mu$m)	&			&			\\ \hline
Y	& 	$0.9$ to $1.1$		&	437		&	1864		\\
J 	& 	$1.2$ to $1.35$	&	265		&	1253		\\
H 	& 	$1.5$ to $1.65$	&	94		&	1175		\\
K 	& 	$2.0$ to $2.4$		&	71		&	774		\\
L 	& 	$3.0$ to $4.0$		&	18		&	15		\\
M 	& 	$4.5$ to $5.3$		&	4		&	0		\\ \hline

\end{tabular}
\end{center}
\label{tab:kerber}
{\bf Notes.} A comparison between the number of thorium lines found by \cite{kerber2008th} and the number of uranium lines found in this work.  Lines without identifications from this work (indicated with a ``{\bf ?}'' symbol in the line list) are not accounted for in this table.  Note that this is not a comparison between the true number of lines that these two sources provide in these regions, but rather a snapshot of the current number of known lines. 
\end{table}

\section{Conclusions}
\label{sec:conclusions}

We have measured \nlines uranium lines in the NIR, which represents a significant increase in the number of calibration lines over the current NIR emission standard, Th/Ar.  This line list has been generated specifically with astronomical applications in mind --- we have taken care to avoid blended lines, and where we could not do so, we have indicated suspected or likely blends in the table.  We anticipate this line list will be useful on several upcoming astronomical spectrographs, including APOGEE \citep{allende2008apogee}, CARMENES \citep{2010SPIE.7735E..37Q}, and iSHELL \citep{2008SPIE.7014E.208T}.  

\section{Acknowledgements}

We acknowledge support from NASA through grant NNXlOAN93G and the NAI and Origins grant NNX09AB34G, and the NSF grant AST 1006676. Financial support is provided by NSF under Grant AST-0906034 and NIST.  This research was partially performed while SLR held a National Research Council Research Associateship Award at NIST.  This work was partially supported by funding from the Center for Exoplanets and Habitable Worlds. The Center for Exoplanets and Habitable Worlds is supported by the Pennsylvania State University, the Eberly College of Science, and the Pennsylvania Space Grant Consortium.  We would also like to thank Joe Reader, John Curry, Florian Kerber, and the anonymous referee for carefully reading the paper and providing useful feedback.

\bibliographystyle{apj}

\end{document}